\documentclass[referee]{aa}
\usepackage{graphicx,natbib}
\bibpunct{(}{)}{;}{a}{}{,}
\begin{document}

\title{Near-infrared spectra of ISO selected Chamaeleon I young stellar
objects\thanks{Based on observations collected at the European Southern
Observatory, Chile, (ESO proposal N.65.I-0054)}}                    

%\subtitle{}

\author{Mercedes G\'omez\inst{1}\and Paolo Persi\inst{2}}
\institute{Observatorio Astron\'omico de C\'ordoba, Laprida 854, 5000
C\'ordoba, Argentina \\ email: mercedes@oac.uncor.edu\and
Istituto Astrofisica Spaziale e Fisica Cosmica, CNR, 
Via del Fosso del Cavaliere, 100, 00133, Roma, Italia}

\offprints{M. G\'omez}
                                                                                                           
\date{Received January 15, 2002; accepted April 16, 2002}

\abstract{We present 0.95--2.5 $\mu$m moderate (R $\sim$ 500)
resolution spectra of 19 ISOCAM detected sources in the Chamaeleon I dark cloud. 
Thirteen of these stars are candidate very low mass members of the cloud 
proposed by \citet{per00} on basis of the mid-IR color excess.
The sample also includes a bona-fide young brown dwarf (Cha H$\alpha$ 1), 
a {\it transition} ---stellar/sub-stellar--- object \hbox{(Cha H$\alpha$ 2),}
one previously known T Tauri star (Sz 33) and three ISOCAM sources with
no mid-IR excess.  The spectra of the mid-IR color excess sources are relatively flat
and featureless in this wavelength range. Both atomic and molecular
lines (when in absorption) are partially veiled suggesting the presence
of continuum emission from circumstellar dust. In addition some of the
sources show Paschen and Brackett lines in emission.  We apply the
2 $\mu$m water vapor index defined by \citet{wil99} to estimate
spectral types.  These stars have spectral
types M0--8. We use Persi et al.'s stellar luminosity determinations,
in combination with D'Antona \& Mazzitelli latest pre-main sequence
evolutionary tracks, to estimate masses and ages.
The ISOCAM detected mid-IR excess sources have sub-solar masses
down to the H-burning limit and a median age of few $\times$10$^6$ yr,
in good agreement with the higher mass members of this cloud.
}

\authorrunning{G\'omez and Persi}
\titlerunning{ISO selected Cha I young stellar objects}

\maketitle
%
%________________________________________________________________

\section{Introduction}

Recently \citet{per00} have presented the results of the
ISOCAM survey of the Chamaeleon I dark cloud. These observations
cover $\sim$ 0.59 deg$^2$ of the cloud at 6.7 and 14.3 $\mu$m 
to a detection limit of about 0.7 mJy and 1.3 mJy, respectively.
They detected 282 sources at 6.7 $\mu$m; 103 of which were also
found at 14.3 $\mu$m, including 74 previously known members of the cloud.
These data, in combination with the DENIS near-IR observations \citep{cam98},
have allowed \citet{per00} to identify 34 very low
luminosity (L $\sim$ 0.02--0.3 L\sun) candidate
members of the Chamaeleon I dark cloud. 
These newly detected objects show significant mid-IR excess and fluxes 
at 6.7 $\mu$m $>$ 2.5 mJy. They  probably include very low mass stars in the cloud
down to (and even below) the H-burning limit.  We take advantage of the recent
development of the near-IR spectroscopy to unveil the physical
nature of thirteen of the ISOCAM proposed low mass members of
the Chamaeleon I dark cloud. Most of our targets
have $K$ $\sim$ 11 (see Table \ref{ISOSP}) and no
optical counterparts on the Digitized Sky Survey.
Thus the near-IR wavelengths provide a suitable range
to study these sources.     

$K$-band spectroscopic observations of young stellar objects
in different star-forming regions at similar spectral
resolution have been reported by several authors \citep{su89, car89,
car90, cama92, caei96, grme95}.
In addition, \citet{grla96} have published
an extensive atlas of near-IR spectra of about 100 young stellar objects.
\citet{wil99}, \citet{cus00}, and \citet{luc01} have used near-IR spectra to study
candidates and bona-fide young brown dwarfs in the $\rho$ Ophiuchi and 
Orion clouds.  These works provide a detailed reference to our
targets as well as initial determinations of spectral features and
characteristics usually present in the spectra of young stellar
objects of different spectral energy distribution classes 
and masses. 

We describe our observations and data reduction in \S2.
In \S 3 we analyze the spectral features of these sources
and compare them with the near-IR spectroscopic characteristics of
previously known young stellar objects.  We apply the 2 $\mu$m water vapor
index defined by \citet{wil99} to determine
spectral types. We use \citet{per00}'s stellar luminosity determinations,
in combination with  D'Antona \& Mazzitelli latest pre-main sequence evolutionary
tracks, to estimate masses and ages for these stars.  We discuss our results
and conclude with a brief summary in \S 4.             
                                                        
%__________________________________________________________________

\section{Observations and data reduction}

We obtained 0.95--2.5 $\mu$m spectra of 19 sources in the
Chamaeleon I dark cloud. Thirteen of these sources are ISOCAM
detected candidate young stars with significant mid-IR color excess.
Three are previously known young members of the cloud (Cha H$\alpha$ 1, 
Cha H$\alpha$ 2, and Sz 33) and other three are ISOCAM detected sources
with no mid-IR excess. The 13 targets with mid-IR color excess
were selected {\it randomly} among the 34 ISOCAM candidates proposed by
\cite{per00}. Table \ref{ISOSP} reports our sample, the infrared magnitudes 
and some derived photometric parameters obtained from the literature.
We also observed three previously known pre-main sequence stars 
(Sz 84, Sz 97, and Sz 105) belonging to the Lupus star-forming region.
These known stars have M spectral types, optically determined, and
will be used as comparison and reference to the ISOCAM sources. 

The observations were carried out on April 27--28 2000 with the ESO NTT
near-IR spectrograph/imaging camera SOFI (Son OF ISAAC).
We observed our targets with the two low resolution grisms (red and blue)
to roughly cover the $JHK$ bands on a Hawaii HgCdTe 1024$\times$1024
detector at a plate scale of 0.292$''$/pix.
The blue grism covers the spectral region between
0.95--1.63 $\mu$m and the red grism the region between
1.53--2.52 $\mu$m. The corresponding spectral resolutions
(R $=$ $\lambda$/$\Delta$$\lambda$) are 930 and 980 for a 0.6$''$ slit. 
We used a 1$''$ width and 290$''$ length
slit. This slit width provides an effective spectral resolution
of approximately 560 and 590, blue and red grisms.
The corresponding dispersions are 11.5 \AA/pix and 17.2 \AA/pix, respectively. 

\begin{table}
\caption[]{Compiled photometric magnitudes and derived parameters for the 
nineteen ISO sources spectroscopically observed} \label{ISOSP}
\begin{tabular}{lrrcrrrrllll}
Name     &  $I^{\mathrm{a}}$  & $K^{\mathrm{b}}$   & $H-K^{\mathrm{b}}$ &
$J-H^{\mathrm{b}}$ &
$L^{\mathrm{c}}$   & mag6.7$^{\mathrm{d}}$  & mag14.3$^{\mathrm{d}}$ &
A$_\mathrm{J}$$^{\mathrm{d}}$ &
L/L\sun$^{\mathrm{d}}$ & $\alpha_{\mathrm{IR}}$$^{\mathrm{d}}$ & Other ID \\
   \hline
   \noalign{\smallskip}
 ISO-ChaI 79  &       & 12.39 &  1.23 & 1.78  &       & 9.83   &  6.39   & 2.4$^*$ & 0.07 & $-$0.8 & \\
 ISO-ChaI 95  & 16.58 & 12.31 &  0.45 & 0.73  &       & 10.03  &  6.20   & 0.3 & 0.02 &        & Cha H$\alpha$ 1$^{\mathrm{e}}$\\
 ISO-ChaI 98  & 17.61 & 11.89 &  0.71 & 1.38  &       & 9.45   &  6.47   & 2.0 & 0.07 & $-$1.2 & \\
 ISO-ChaI 111 & 15.26 & 10.65 &  0.57 & 1.09  & 9.94  & 8.62   &  5.46   & 1.3 & 0.08 &        & Cha H$\alpha$ 2$^{\mathrm{e}}$\\
 ISO-ChaI 138 & 16.74 & 13.05 &  0.41 & 0.69  &       & 9.13   &  5.77   & 1.3 & 0.03 & $-$0.2 & \\
 ISO-ChaI 143 & 15.43 & 11.09 &  0.55 & 1.09  & 10.67 & 9.11   &  5.84   & 1.3 & 0.11 & $-$1.2 & \\
 ISO-ChaI 154 & 14.42 & 11.37 &  0.33 & 1.01  &       & 10.22  &         & 0.3 & 0.04 &        & \\
 ISO-ChaI 158 & 13.56 & 11.25 &  0.29 & 0.70  &       & 11.34  &         &     &      &        & \\
 ISO-ChaI 209 &       & 12.22 &  1.24 & 2.00  &       & 10.03  &  6.77   & 3.0$^*$ & 0.06 & $-$1.2 & DENIS-P J1109.8-7714$^{\mathrm{a}}$ \\
 ISO-ChaI 220 & 18.11 & 12.48 &  0.85 & 1.25  &       & 10.28  &  6.82   & 2.0 & 0.05 & $-$1.1 &  \\
 ISO-ChaI 224 & 13.99 &  9.10 &  0.74 & 1.35  & 8.22  & 7.45   &  4.10   & 1.4 & 0.48 &        & Sz 33$^{\mathrm{f}}$\\
 ISO-ChaI 225 & 17.12 & 12.75 &  0.98 & 1.11  &       &  9.08  &  5.81   & 0.8 & 0.02 & $-$0.2&  \\
 ISO-ChaI 235 & 17.93 & 11.18 &  0.90 & 1.49  &       & 10.94  &         & 2.3 & 0.14 & $-$1.1&  \\
 ISO-ChaI 238 & 17.58 & 12.19 &  0.83 & 1.42  &       &  9.10  &  5.79   &     & &       &  \\
 ISO-ChaI 239 & 15.45 & 10.47 &  0.56 & 1.47  & 9.85  &  9.85  &         &     & &       &  \\
 ISO-ChaI 250 & 15.57 & 10.71 &  0.60 & 1.52  &       &  8.84  &  7.03   & 1.6 & 0.20 & $-$1.9&  \\
 ISO-ChaI 252 & 17.11 & 12.29 &  0.60 & 1.12  &       & 10.24  &  6.80   & 1.7 & 0.06 & $-$1.1&  \\
 ISO-ChaI 256 & 17.51 & 10.78 &  0.93 & 1.62  &       &  8.13  &  4.69   & 2.4 & 0.17 & $-$0.8&  \\
 ISO-ChaI 282 & 15.52 & 11.88 &  0.61 & 1.00  &       &  9.89  &  6.60   & 1.0 & 0.06 & $-$1.4&  \\
 \noalign{\smallskip}
 \hline
\end{tabular}
\begin{list}{}{}
\item[$^{\mathrm{a}}$\citet{cam98}]
\item[$^{\mathrm{b}}$\citet{goke01}]
\item[$^{\mathrm{c}}$\citet{kego01}]
\item[$^{\mathrm{d}}$\citet{per00}]
\item[$^{\mathrm{e}}$\citet{com98}]
\item[$^{\mathrm{f}}$\citet{sch77}]
\end{list}                                  

\noindent
$^*$Note: Calculated from A$_\mathrm{J}$ $=$ 2.63[($J-H$)$-$($J-H$)$_\mathrm{o}$],
adopting ($J-H$)$_\mathrm{o}$$=$ 0.85 \citep[see][]{per01}.

\noindent 
\end{table}

The integration time and the number of exposures per target
were chosen according to the brightness of the individual sources
and background contribution during each night.  In most 
cases we obtained a total of 4 blue spectra of the same 
integration time, and 6 red spectra.
Typical integration times are 80--400 sec for the blue
and 180--240 sec for red grisms, respectively.  The telescope was
nodded 30--60$''$ along the slit between consecutive positions
following the usual ABBA pattern. This procedure corresponds to the
``Nod Throw Along Slit'' scheme as fully described in the SOFI Users Manual
\citep{lid00}.

In addition to the program sources we observed several
atmospheric standards from the SOFI list of infrared spectroscopic
standards (see the SOFI web
page\footnote{http://www.ls.eso.org/lasilla/Telescopes/NEWNTT/sofi/index.html})
for telluric absorption corrections.
We selected two groups of standards with similar airmass as our candidate
sources and observed them periodically (i.e., every $\sim$ 2 hours) during
each night.  These telluric stars comprise both late 
(G3-5) and early (O8-9) spectral type objects.  We took 4 spectra per
observations in both spectral regions. Total integration times ranged
between 8 and 20 sec for the two grisms. The stars were shifted along
the slit direction by 60$''$ between exposures, following the ABBA pattern.

We obtained multiple flat field images, in both grisms,
with a dome screen, using incandescent lamps on and off.
A xenon lamp, also taken on and off, each night provided the wavelength
calibration for our data. 

To reduce the data we used IRAF\footnote{IRAF is distributed by
the National Optical Astronomy Observatory, which is operated by
the Association of Universities for Research in Astronomy, Inc.
under contract to the National Science Foundation.}. We subtracted
one image from another (using pairs of nodded observations) to
eliminate the background and sky contribution in first approximation.
This subtraction automatically
took care of dark current and bias level.  We chose the images closer
in time and position on the sky to perform this subtraction. We
flat-fielded our data dividing by an appropriated normalized
flat-field for each of the grisms. The flat-fields were
created median-filtering multiple exposures in each grism. 
Bad pixels were replaced by linear interpolation from neighboring
pixels. 

We aligned individual exposures corresponding to each grism and
target and combined the corresponding frames
into one blue and one red image per object.
To do this alignment as accurate as possible we traced each
spectrum as describe below and determine a precise
center.  Then we shifted all data, corresponding to each target and grism,
to a common position and co-added them.

We used the {\it twodspec} task APALL to trace and extract the
spectra along a 12 pixel wide aperture on the co-added
blue and red images. A further sky subtraction was
done by fitting a polynomial to the regions on either
side of the aperture.  Residual sky contributions
were corrected reasonably well by this procedure. 
A non linear low order fit to the lines in the xenon lamp
was used to wavelength calibrate the spectra. 
We verified our wavelength solution using the positions of
well known telluric lines as well as some H lines in the spectra of the standard
stars. 

We removed telluric features from our data dividing the program
spectra by the atmospheric standard spectra.
At the spectral resolution used (R $\sim$ 500--600) the O8-9
stars provided quasi featureless spectra in all the spectral
range covered by the two grisms.  The only spectral lines present
are strong H and He lines. We used the {\it onedspec} task SPLOT to
remove H (Paschen and Brackett)  and He (I and II) lines from
the spectra of the O stars, interpolating across each line. G5-3 stars, 
on the contrary, display weak Paschen and Brackett series but
contain atomic (metallic) lines in addition to some molecular
features.  The spectra of these solar type stars were used with not
modifications (i.e., no lines were removed). 

For our ISOCAM targets we obtained two telluric corrected
spectra using as telluric calibrators the early and the
solar type atmospheric standard spectra.  These standards
differed from the science targets by $<$ 0.2 in airmass.
In addition, the time interval between the observations
of the pair target-standard was typically $<$ 1 hour. 

The division by the O8-9 standard canceled out
telluric features in the science spectrum reasonable well but
introduced the inverse of the atmospheric standard in
the corrected spectrum.  We recovered the true spectral shape
multiplying the resultant spectrum by a Planck function at
the temperature corresponding to the atmospheric star.
We also used the atmosphere models of Kurucz \footnote{Available at
http://cfaku5.harvard.edu/grids.html.} and the library of spectra
from \citet{pic98} for the appropriate T$_\mathrm{eff}$.  These
procedures gave essentially the same result. 

We used the normalized solar spectrum \footnote{NSO/Kitt Peak
FTS data used here were produced by NSF/NOAO.},
convolved to the SOFI blue and red grism resolutions, to multiply 
the second set of science spectra, the ones divided
by the G3-5 telluric standards. We used a Gaussian profile
to smooth the solar spectrum and a Planck function
at the temperature of the telluric standard to reproduce the
corresponding stellar continuum. Kurucz models and Pickles 
spectra for the telluric continuum, combined with the appropriate convolved
normalized solar spectrum, produced basically identical corrected spectra 
for our science targets.

We finally combined the blue and the red grism spectra for each
object and eliminated regions of deep atmospheric absorptions from our
analysis as not satisfactory corrections were obtained in these
regions. In addition we also trimmed out wavelengths between 0.95--0.978 $\mu$m 
as edge effects and a poor S/N ratio were present.
The useful spectral ranges are: 9780--11000 \AA, 11600--13300 \AA,
14600--17800 \AA, and 20500--25000 \AA.  The resultant telluric corrected spectra
using both set of standards were essentially identical and display the spectral
shapes and features of the science targets.  In Sect. 3.2
we chose to show the set of science spectra obtained applying the
G3-5 standards as telluric corrector. 

\section{Data analysis and results}

\subsection{The near-IR $J-H$ vs. $H-K$ diagram}

\begin{figure}
\centering
\includegraphics[width=14cm]{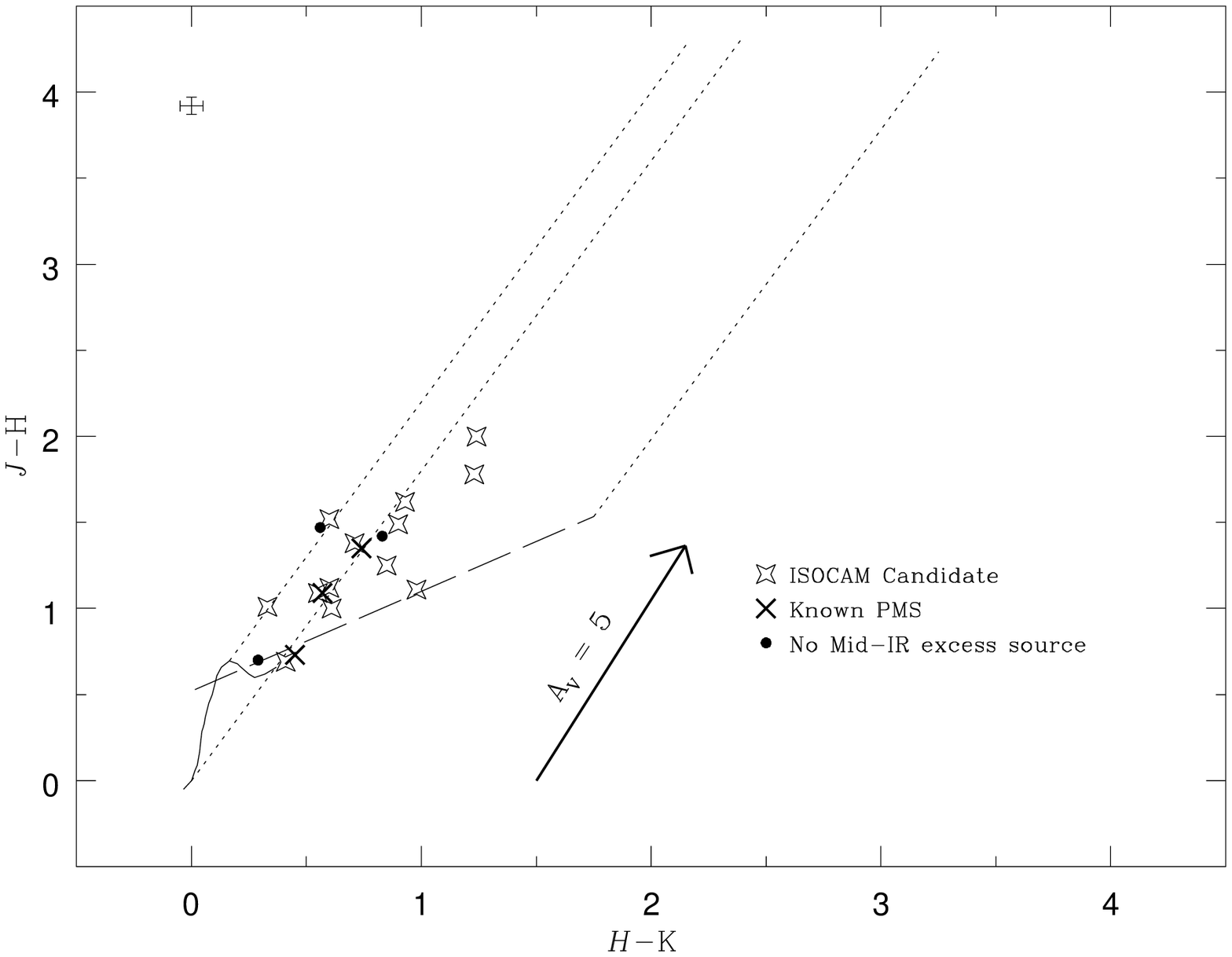}

\vskip -1.0in

\caption{Near-IR color-color diagram for the ISOCAM sources listed
in Table \ref{ISOSP}.  The solid and long-dashed lines indicate the loci of unreddened main
sequence dwarfs \citep{bebr88} and CTTS \citep{mey97}. The
dotted lines define the reddening band, corresponding to a
reddening vector E($J-H$)/E($H-K$) $=$ 1.80 \citep{goke01}. 
The arrow indicates an A$_\mathrm{V}$ $=$ 5 mag \citep{rile85}. Typical photometric
errors are displayed in the upper left corner. }
\label{Fig1}
\end{figure}

Table \ref{ISOSP} compiles photometry and derived parameters 
for the 19 sources reported in this paper from the literature. 
Fig. \ref{Fig1} shows the positions of these stars in
the near-IR color-color plot. The dotted symbols and the crosses
represent stars in our sample with no mid-IR excess and
previously identified members of the cloud, respectively. The big
stars correspond to objects with mid-IR excess.  Superposed on this
diagram are the loci of unreddened main sequence dwarfs
\citep[solid line,][]{bebr88} and CTTS --classical T Tauri stars-- 
\citep[long-dashed line,][]{mey97}. The reddening band \citep[dotted line,][]{goke01}
and the length of the reddening vector \citep{rile85} are also indicated. 

Taking into account typical photometric errors for these sources
\citep[$\sim$ 0.03 mag for \hbox{$K =$ 11,} see ][]{goke01},
we found that approximately 70\% (9 out of 13) of the stars selected on
basis of the mid-IR excess show no significant near-IR excess emission.
\citet{per00} classified these ISOCAM detected stars as
Class II members of the Chamaeleon I dark cloud using the infrared
spectral index ($\alpha_{\mathrm{IR}}$ $=$ dlog($\lambda$F$_{\lambda}$/dlog($\lambda$)), 
computed from 2.2 to 14.3 $\mu$m. 
They typically have $-$1.6 $<$ $\alpha_{\mathrm{IR}}$ $<$ 0.3, corresponding to Class II
objects (see Table \ref{ISOSP}). The only exception is ISO-ChaI 250 with $\alpha_{\mathrm{IR}}$ $=$ $-$1.9, 
identified as a Class II-III member of the cloud by \citet{per00}.  The CTTS 
or the Class II objects usually have
detectable near-IR excesses. However, a significant fraction 
of the bona-fide CTTS in the Taurus \cite[e.g.,][]{str93,keha95} and in the
\hbox{Chamaeleon I} \citep[e.g.,][]{gast92,com00} star forming regions shows no measurable near-IR
excesses. Therefore, mid-IR observations are a very powerful tool
to discriminate young stellar objects from background stars  
\citep[cf.][]{per00}. 

\subsection{The near-IR spectra}

Figs. \ref{Fig2} -- \ref{Fig7} show the infrared
spectra of the 19 sources reported in this paper.  Prominent atomic and molecular
features in the 0.98--2.45 $\mu$m range for late type stars (i.e., spectral types later than G0) 
are labeled \citep{wal00, mey98, wahi97}.  Table \ref{ISOEW} lists equivalent widths 
for the some of the strongest lines. We estimate an uncertainty of $\sim$ 1-2 \AA~ in the 
Pa$\beta$, Br$\gamma$,  Na I doublet (2.206 and 2.209 $\mu$m), and Ca I triplet
(2.261, 2,263, and 2.266 $\mu$m) equivalent widths. For the combine CO $\nu$ $=$ 0-2 and 2-4 bands
the uncertainty in our measurements is $\sim$ 3 \AA.  Stars in Table \ref{ISOEW} have, on average,
similar or slightly smaller equivalent widths than M type standards \citep{klha86,wal00}, with 
exception of few sources (such as ISO-ChaI 79, ISO-ChaI 143, ISO-ChaI 225, etc.) with
no absorption lines present at the spectral resolution used. 

\begin{table}
\caption[]{Equivalent widths (in \AA) for the nineteen ISO sources analyzed$^*$} \label{ISOEW}
\begin{tabular}{lrrrrr}
Name     & Pa$\beta$    &  Br$\gamma$  & Na I$^{\mathrm{a}}$ & Ca I$^{\mathrm{b}}$         & CO$^{\mathrm{c}}$ \\ 
         & 1.28$\mu$m &  2.17$\mu$m & 2.21 $\mu$m & 2.26 $\mu$m & 2.23--2.38 $\mu$m \\
\hline
\noalign{\smallskip}
 ISO-ChaI 79  &        &       &    &      &    \\
 ISO-ChaI 95  &     0.7&    4.5& 4.8&   2.0& 37.5\\ 
 ISO-ChaI 98  &  $-$2.9& $-$3.6&    &   3.0& 16.5\\
 ISO-ChaI 111 &     1.4&    3.9&    &      & 11.9\\
 ISO-ChaI 138 &  $-$1.5& $-$0.7& 4.7&   3.4& 27.2\\
 ISO-ChaI 143 &  $-$3.2&       &    &      &     \\
 ISO-ChaI 154 &     2.3&       &    &      &     \\
 ISO-ChaI 158 &     1.9&    2.9&    &      & 13.7\\
 ISO-ChaI 209 &     5.1&   11.2&    &      & 31.7\\
 ISO-ChaI 220 &     2.9& $-$4.6&    &      &  5.5\\
 ISO-ChaI 224 &  $-$3.0& $-$1.5& 2.5&   2.3& 18.9\\
 ISO-ChaI 225 &        &       &    &      &     \\ 
 ISO-ChaI 235 &     1.3&    5.0&    &   4.5&  20.0\\
 ISO-ChaI 238 &        &       &    &      &  40.5\\
 ISO-ChaI 239 &     2.0&    3.2& 2.3&   1.4&  43.8\\
 ISO-ChaI 250 &     3.2&       & 4.8&   4.8&  34.6\\
 ISO-ChaI 252 & $-$27.2&       & 7.7&      &  31.7\\
 ISO-ChaI 256 &  $-$5.6&       & 2.4&   4.1&  28.0\\
 ISO-ChaI 282 &     0.9&       &    &      &  37.0\\
 \noalign{\smallskip}
 \hline
\end{tabular}
\begin{list}{}{}
\item[$^{\mathrm{a}}$ Na I doublet (2.206 and 2.209 $\mu$m)]
 
\item[$^{\mathrm{b}}$ Ca I triplet (2.261, 2,263, and 2.266 $\mu$m)]
 
\item[$^{\mathrm{c}}$ CO $\nu$ $=$ 0-2 and 2-4 bands]
\end{list}
\noindent
$^*$Note: Positive values indicate absorptions and negative values correspond
to emissions. Missing equivalent widths indicate that the corresponding feature
was not detected either in absorption or emission at the resolution used
(R $\sim$ 950).

\noindent
\end{table}                                                   

\subsubsection{ISO detected sources with no mid-IR excess}

\begin{figure}
\centering
\includegraphics[width=17cm]{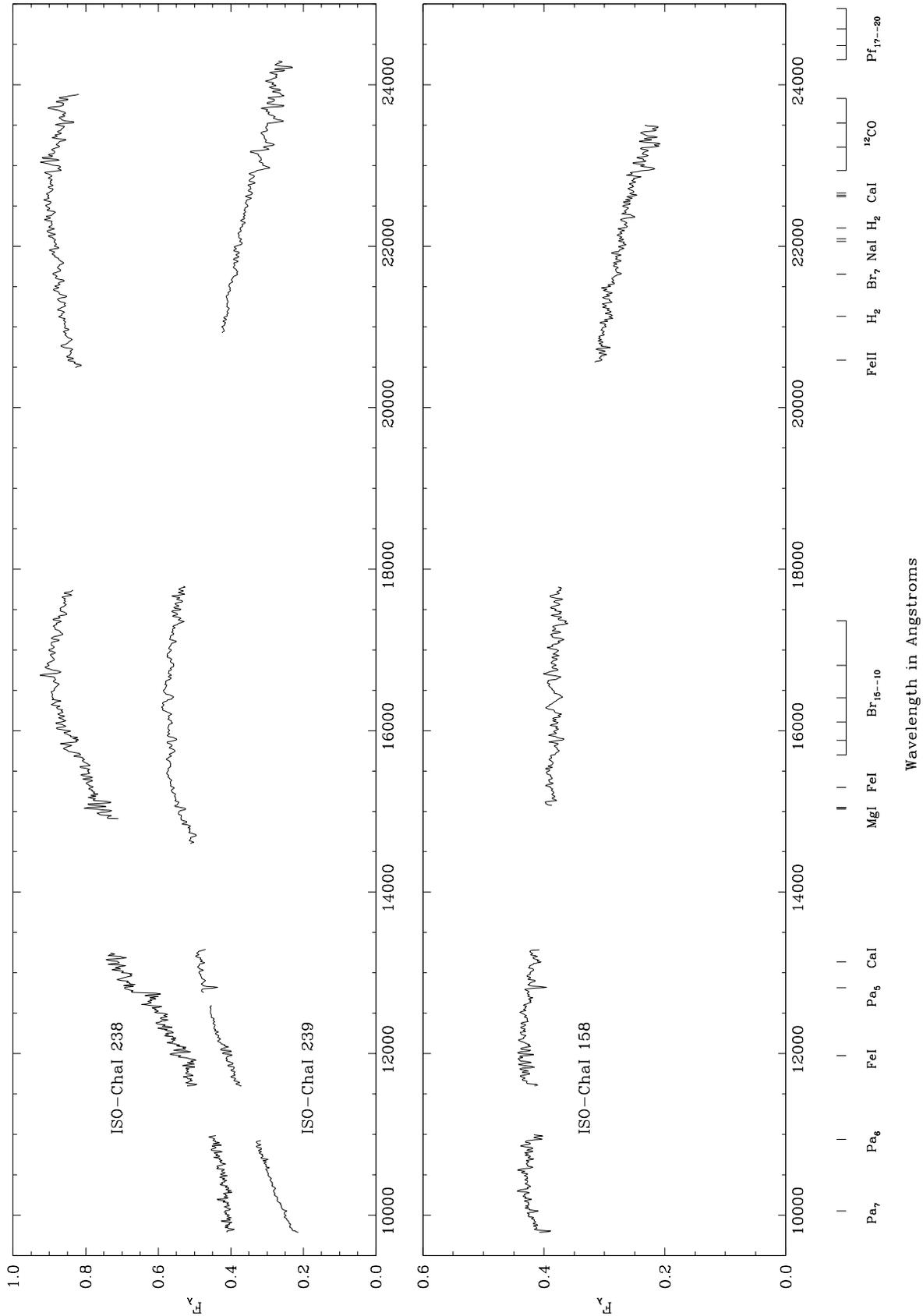}
\caption{$JHK$ band spectra of ISOCAM detected sources in
the Chamaeleon I dark cloud with no mid-IR excess. The vertical
scale (F$_{\lambda}$) corresponding to ISO-ChaI 238 is offset
by 0.82 with respect to ISO-ChaI 239.
Prominent atomic and molecular features in the
0.98--2.45 $\mu$m range are labeled.}
\label{Fig2}
\end{figure}

Near-IR spectra of three ISOCAM detected sources without near- and mid-IR
excesses (ISO-ChaI 158, ISO-ChaI 238, and ISO-ChaI 239) are illustrated in 
Fig. \ref{Fig2}. The spectra of these targets show no
sign of activity, such as emission lines (at the spectral resolution used),
characteristic of young stars. As we will derive in Sect. 3.3.1 these stars
have M type spectral types and are likely background objects, seen at
the near-IR wavelengths through the cloud. 

\subsubsection{Previously known young stellar objects}

\begin{figure}
\centering
\sidecaption
\includegraphics[width=17cm]{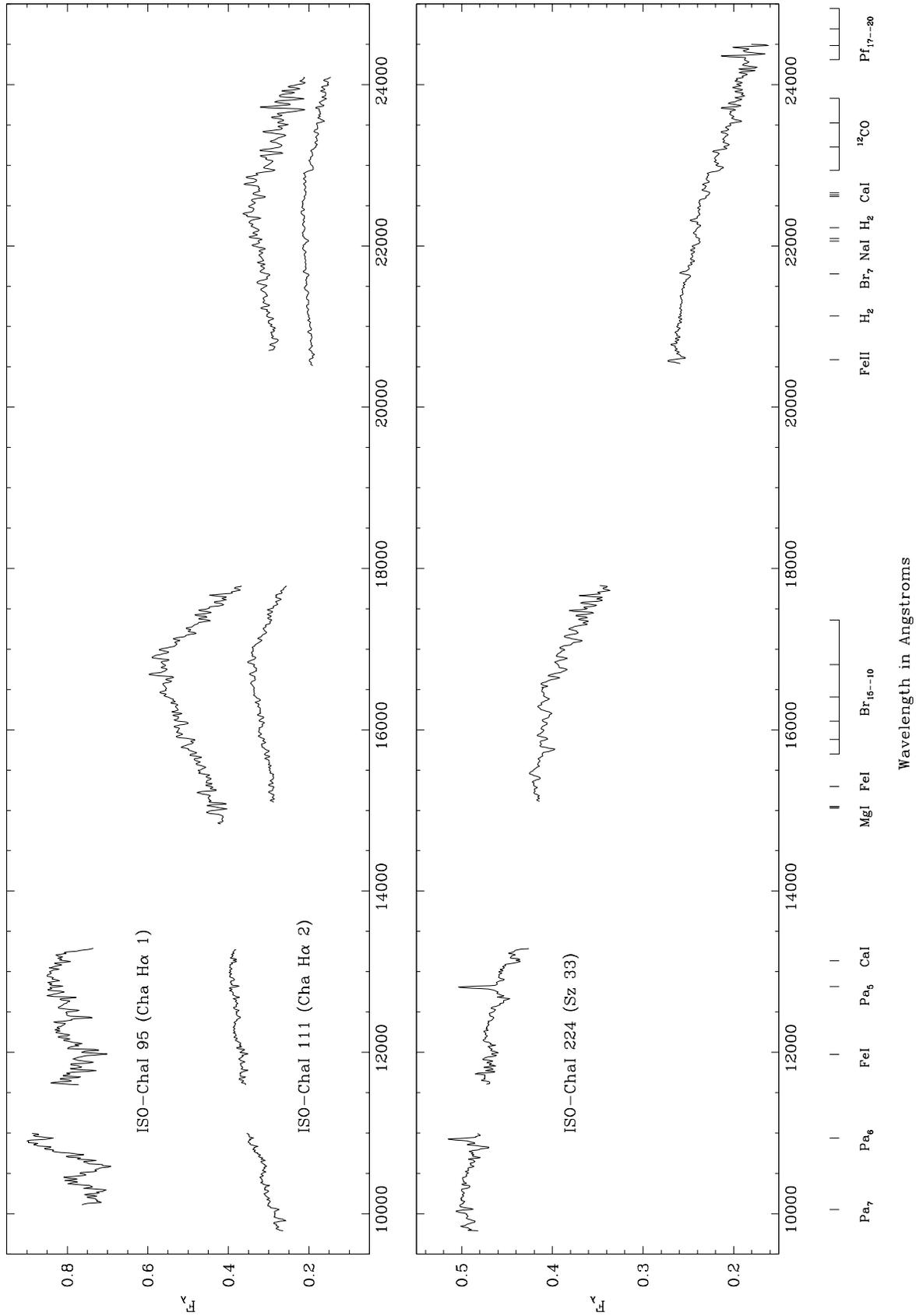}
\caption{$JHK$ band spectra of three previously known young stellar
objects in the \hbox{Chamaeleon I} dark cloud.  
Prominent atomic and molecular features in the
0.98--2.45 $\mu$m range are labeled.}
\label{Fig3}
\end{figure}

Fig. \ref{Fig3} shows our near-IR spectra of three previously known
young stellar objects in the Chamaeleon I dark cloud: ISO-ChaI 95 (Cha H$\alpha$ 1),
ISO-ChaI 111 (Cha H$\alpha$ 2), ISO-ChaI 224 (Sz 33).
Cha H$\alpha$ 1 (M $\sim$ 0.05 M\sun) and Cha H$\alpha$ 2
(M $\sim$ 0.08 M\sun) are a bona-fide brown dwarf and a {\it transition}
(stellar/sub-stellar) object of the cloud
\citep{com00}.  \citet{com98} measured EW(H$\alpha$) of 59 and 39 \AA~ for 
Cha H$\alpha$ 1 and \hbox{Cha H$\alpha$ 2,} respectively, corresponding to 
the EW(H$\alpha$) associated with the solar mass classical T Tauri
stars (CTTS). They also estimated optical 
M7.5-8 and M6 spectral types for these sub-stellar or quasi sub-stellar
objects.  \citet{neco99} detected Li 6707 \AA~ in higher 
resolution spectra assuring the pre-main sequence status
of these objects.

\citet{com00} obtained $HK$ band spectra for a sample of 11 extremely 
low mass members of the Chamaeleon I dark cloud, including Cha H$\alpha$ 1
and Cha H$\alpha$ 2. These objects have spectral types, 
optically determined, within a very narrow range \hbox{(M6--8).} 
These authors published {\it averaged} near-IR spectra for 
stars with basically identically optical spectral types. 
In this manner, the three averaged near-IR spectra 
(representative of objects with M6, M7, and M8 sub-types)
have a better S/N ratio than the individual spectra of each of the 11
sources. Cha H$\alpha$ 1 and Cha H$\alpha$ 2 are included in the M8 and
M6 groups, respectively.
An individual spectrum corresponding to Cha H$\alpha$ 1
was also published by \citet{neco98}.

Our spectra for ISO-ChaI 95 and ISO-ChaI 111
agree well with the {\it averaged} spectra or with the individual
spectrum in the case of ISO-ChaI 95 of previous works.
We additionally observed the $J$ band as shown in Fig. \ref{Fig3}.
Both objects show a decreasing spectral shape toward longer wavelengths.
No prominent spectral features appear in all the wavelength covered.
In particular H (Paschen, and Brackett) lines are not in emission, 
at this spectral resolution.  

\citet{law96} derived a mass of $\sim$ 0.3--0.4 M\sun~ for
Sz 33 \citep[CHX15b=CHXR41;][]{fekr89,fie93}. This sub-solar mass
young star represents the only previously known T Tauri star in 
our sample. It has been classified as a WTTS (weak T Tauri star)
member of the cloud
\citep{gast92,law96}. However, \citet{law96} casted some doubts
upon the WTTS status of this object. In fact, Sz 33
has near-IR spectroscopic characteristics more frequently found
associated with the Class II objects (CTTS) than with the
Class III stars \citep[WTTS; see][]{grla96}. In particular
both Paschen 5 and 6 (Pa$\beta$, Pa$\delta$) and Brackett 7 (Br$\gamma$ )
are in emission
whereas the spectral lines are moderately veiled at our resolution (see Tables
\ref{ISOEW} \& \ref{SP}). The spectral shape of Sz 33 strongly decreases with increasing wavelength. 

\subsubsection{ISO candidate young stellar objects}
 
\begin{figure}
\centering
\includegraphics[width=17cm]{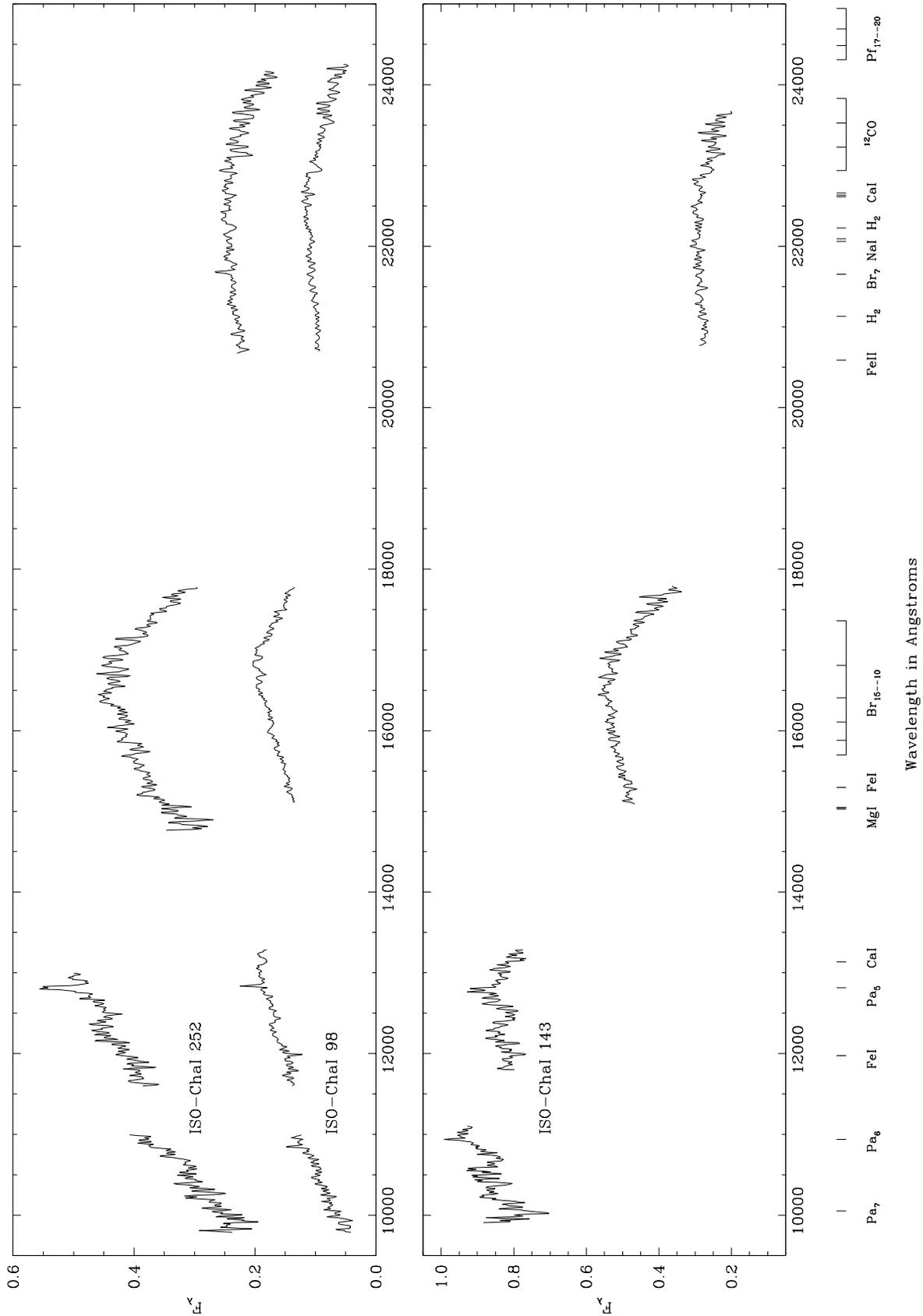}
\caption{$JHK$ band spectra of ISO-ChaI 252, ISO-ChaI 98 and ISO-ChaI 143 in
the \hbox{Chamaeleon I} dark cloud. The vertical scale (F$_{\lambda}$) corresponding
to ISO-ChaI 98 is shifted by $-$0.1 with respect to ISO-ChaI 252. 
Prominent atomic and molecular features in the
0.98--2.45 $\mu$m range are labeled.}
\label{Fig4}
\end{figure}

\begin{figure}
\centering
\includegraphics[width=17cm]{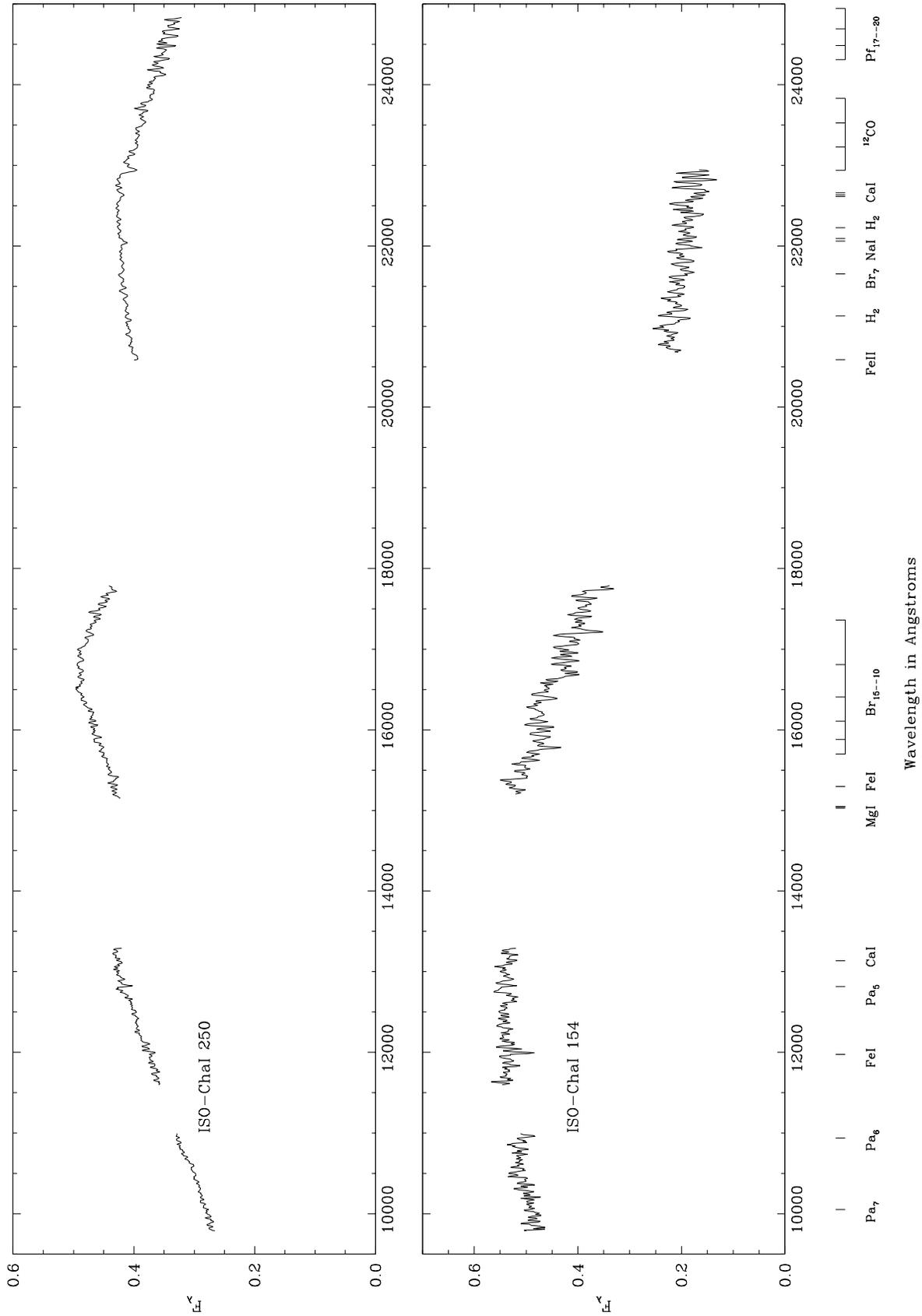}
\caption{$JHK$ band spectra of ISO-ChaI 250, and ISO-ChaI 154 in
the Chamaeleon I dark cloud.  Prominent atomic and molecular features in the
0.98--2.45 $\mu$m range are labeled.}
\label{Fig5}
\end{figure}

\begin{figure}
\centering
\includegraphics[width=17cm]{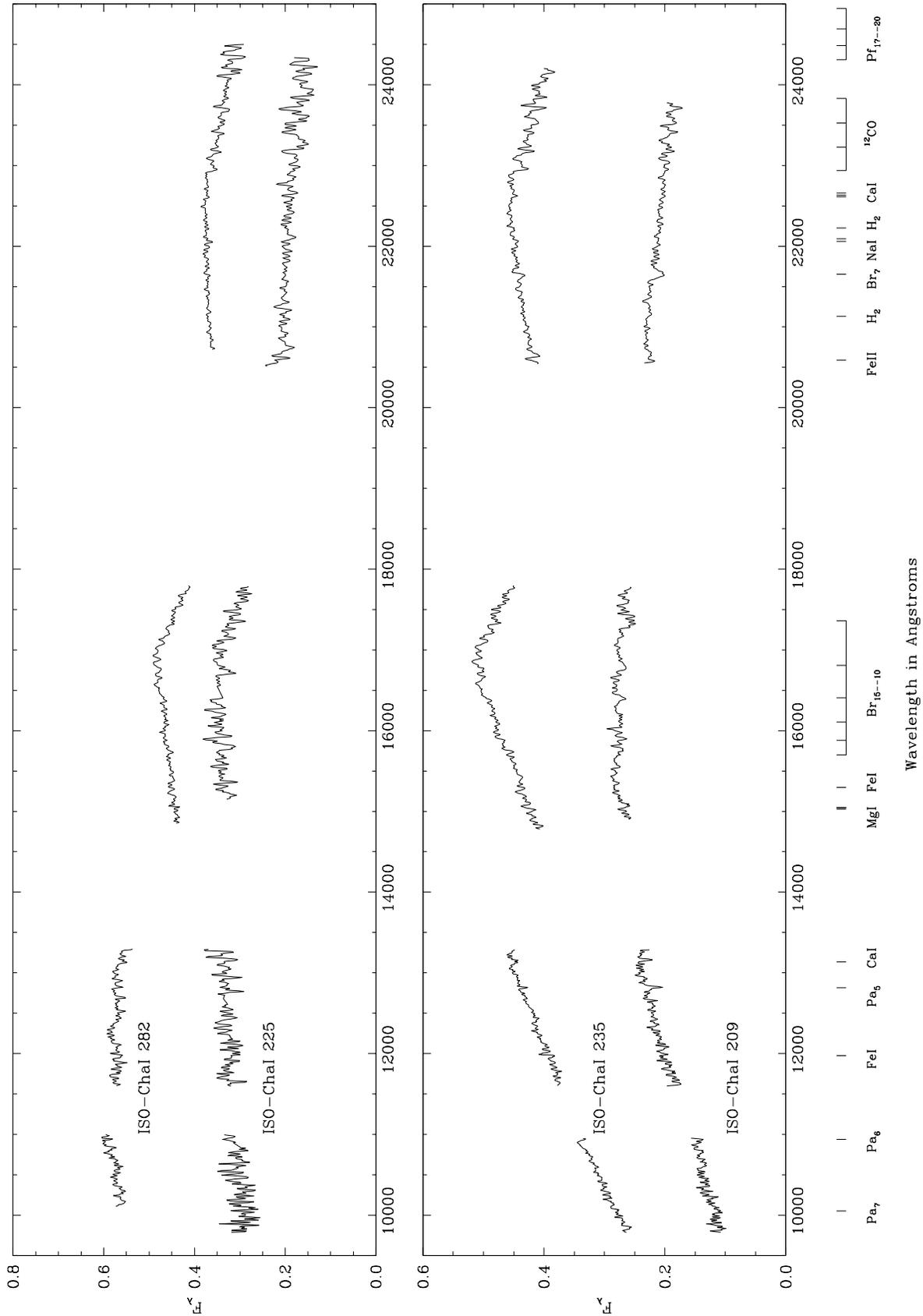}
\caption{$JHK$ band spectra of ISO-ChaI 282, ISO-ChaI 225, ISO-ChaI-235 and ISO-ChaI 209 in
the Chamaeleon I dark cloud.  The vertical scale (F$_{\lambda}$) corresponding
ISO-ChaI 282 (upper panel) is offset by 0.15 with respect to ISO-ChaI 225.
ISO-ChaI-235 (lower panel) is shifted by 0.2 in relation to ISO-ChaI 209.  
Prominent atomic and molecular features in the
0.98--2.45 $\mu$m range are labeled.}
\label{Fig6}
\end{figure}

\begin{figure}
\centering
\includegraphics[width=17cm]{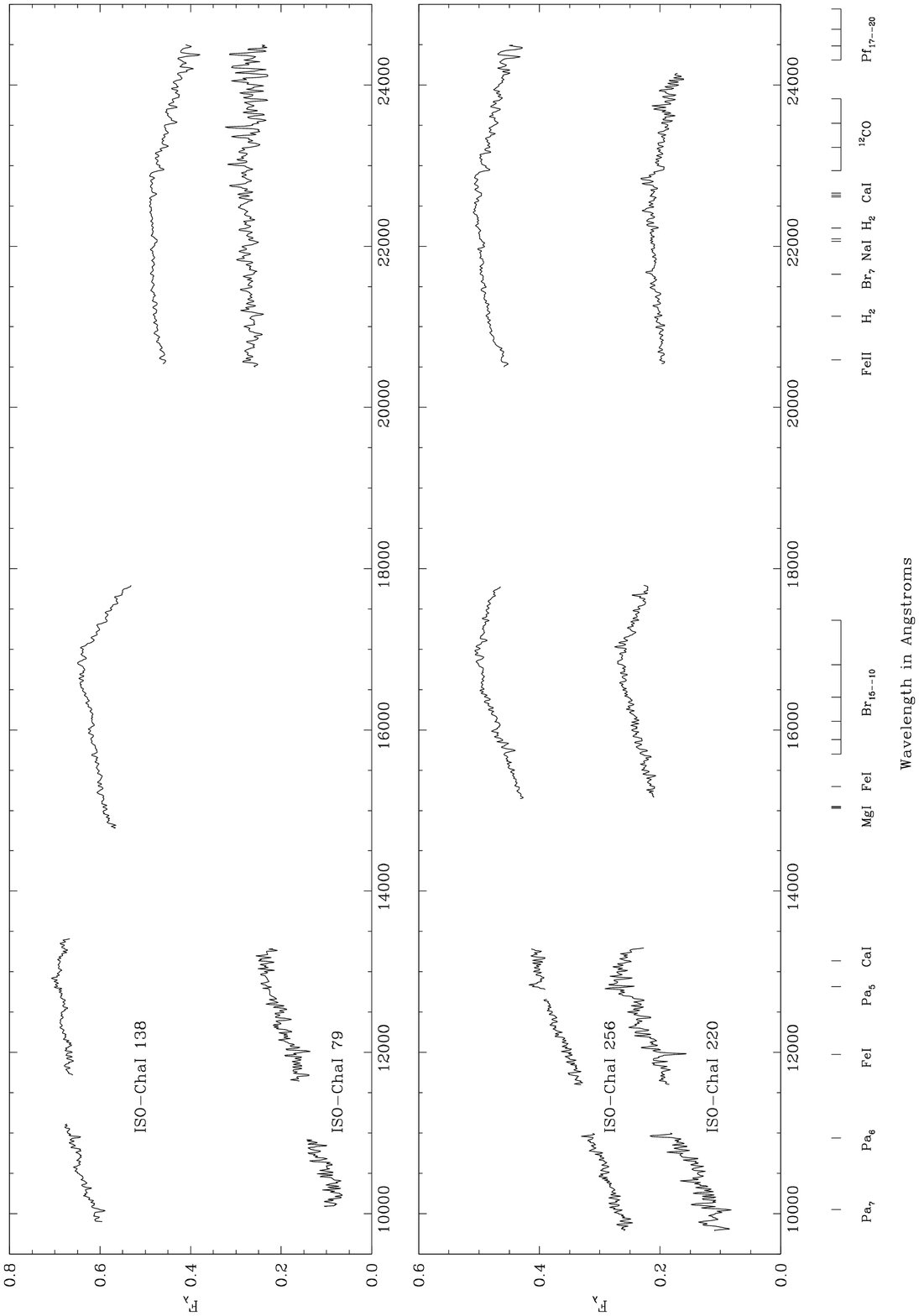}
\caption{$JHK$ band spectra of ISO-ChaI 138, ISO-ChaI 79, ISO-ChaI 256, and ISO-ChaI 220 in
the Chamaeleon I dark cloud.  The vertical scale (F$_{\lambda}$) corresponding to ISO-ChaI 138
(upper panel) is offset by 0.3 with respect to ISO-ChaI 79. ISO-ChaI 256 (lower panel) is
shifted by 0.25 with respect to ISO-ChaI 220. Prominent atomic and molecular features in the
0.98--2.45 $\mu$m range are labeled.}
\label{Fig7}
\end{figure}

Our near-IR spectra for thirteen of the new young stellar objects found by the ISOCAM
survey in the Chamaeleon I dark cloud are
shown in Figs. \ref{Fig4} -- \ref{Fig7}.
As a group these sources show a variety of spectral
shapes. Several stars have fairly constant shapes over the
0.95--2.45 $\mu$m range. Other sources display a smoothly
decreasing spectral shape with increasing wavelength or have a
turnover around 1.5--1.6 $\mu$m. Absorption lines are partially
or complete veiled (see Tables \ref{ISOEW} \& \ref{SP}). ISO-ChaI 79 and
ISO-ChaI 225 show strongly veiled spectra.
These stars, in addition to ISO-ChaI 209 and ISO-ChaI 220, also display the largest near-IR excesses
in Fig. \ref{Fig1} and are thus probably surrounding by significant amount of
circumstellar material. Some objects
(ISO-ChaI 252, ISO-ChaI 98, ISO-ChaI 143, ISO-ChaI 220, and
ISO-ChaI 256, see Table \ref{ISOEW}) have H lines in emission, also indicating  the presence of 
disks as we discuss in Sect. 3.4.

These near-IR spectral characteristics are common among the
Class II population of other well known star-forming regions
such as $\rho$ Ophiuchi and Taurus \citep{grla96}. In general
the Class II objects show a wider range of variation in
different spectral features such as veiling, spectral shape or emission
line intensities. For Class I and Class III objects these
parameters vary over a smaller range or, in other words, the
near-IR spectroscopic characteristics of Class I and \hbox{Class III}
objects are more similar within each group than among
the Class II objects \citep[see]['s atlas]{grla96}.

For ISO-ChaI 79 we have eliminated the spectral region between 13000 \AA~ and
\hbox{20000 \AA~.}  This ISOCAM candidate is one of our faintest targets (see Table
\ref{ISOSP}).  The background contribution was relatively high for this
star making this correction unreliable in this spectral range.  ISO-ChaI 79
was also observed under relatively unfavorable weather conditions resulting in a poor
correction of the deep water atmospheric features and noisy spectrum.
 
ISO-ChaI 154 displays a continuous decreasing spectrum and practically no
evidence of veiling (see Table \ref{SP}) and no emission lines, at the spectral
resolution we used. This spectral behavior is similar to many Class III or WTTS
\citep{grla96}. An optical (4500--8500 \AA) low
resolution spectrum of this object, obtained at the CASLEO (San Juan, Argentina) with the REOSC
spectrograph (DS mode) at the 2.15-m tel., shows no H$\alpha$ in emission. However,
the modest S/N ratio of these data does not allow us to discard the presence
of H$\alpha$ in emission with a relatively small equivalent width 
(i.e., EW(H$\alpha$) $<$ 10 \AA). ISO-ChaI 154 might be a Class III 
member of the cloud.

\subsection{Spectral types and mass determinations for the ISO selected stars}  

\subsubsection{The water vapor index $Q$}  

Figs. 2-7 show that even outside regions of atmospheric water vapor
absorptions, the spectra of these objects have broad stellar H$_2$O absorptions
characteristic of low temperature objects. The strength of these bands 
are extremely sensitive to the spectral types for M dwarfs. Thus the stellar
water features can be used
to estimate spectral types for our targets. \citet{wil99} have derived
the index $Q$ that measures the strength of the H$_2$O bands
around 2 $\mu$m. This index, analogous to the index Q used by \citet{jomo53} 
in the UBV system, is reddening independent. \citet{wil99} 
defined the 2 $\mu$m index $Q$ by the average values of relative flux
density calculated in three narrow bands, F1 (2.07--2.13 $\mu$m), F2(2.267--2.285 $\mu$m),
and F3(2.40--2.50 $\mu$m), as

\begin{equation}
Q=(F1/F2)(F3/F3)^{1.22}.
\label{uno}
\end{equation}

In this definition these authors have adopted the reddening law from
\citet{rile85}, A$_{\lambda}~ \propto~ \lambda^{-1.47}$.
\citet{wil99} applied this index to estimate spectral types for a sample of brown dwarfs
in the $\rho$ Ophiuchi cloud.  The same index, $Q$, has been used by \citet{cus00}
to provide spectral types for additional objects in the same cloud. 

Both \citet{wil99} and \citet{cus00} have obtained linear relations
to determine the M sub-type as function of $Q$ using M dwarf standards with
optically known spectral types \citep[see Eq. (2) in][]{wil99,cus00}. 
These relations have been derived from standards whose
spectra were telluric corrected using AO V stars but whose
continuum shapes were not restored. 

We applied \citet{wil99} and \citet{cus00} calibrations to obtain
spectral types for our targets. We used our atmospheric telluric
divided spectra with no correction for the slope of the
standard spectrum keeping in mind that our atmospheric telluric
standards have G3-5 and O8-9 spectral types and then differ from
those used by these authors.  However, spectral types obtained using both
our sets of standards differ by less than 1.5 sub-types. This
difference is comparable to the accuracy expected by \citet{wil99}
and \citet{cus00} for their calibrations.  As an additional check
we confronted spectral types derived using both our G3-5 and O8-9 spectral 
type standards for a sample of 6 previously known objects.
In Table \ref{KNOWN} we list these stars together with spectral types
derived using our G5 and O8 telluric standards and the optical
spectral types obtained from the literature. Our determinations
agree with the optical types within roughly 1.5 types. 

\begin{table}
\caption[]{Spectral types derived from the $Q$ index using G5 and O8
telluric standards for previously known objects} \label{KNOWN}
\begin{tabular}{lcccl}
Name     &  G5 Tel. Std.  & O8 Tel. Std.  & Optical & Other ID\\
\hline
\noalign{\smallskip}

 Cha H$\alpha$ 1 & M8   & M9   & M7.5$^{\mathrm{a}}$ & ISO-ChaI 95\\
 Cha H$\alpha$ 2 & M5.5 & M6.5 & M6.5$^{\mathrm{a}}$ & ISO-ChaI 111\\
 Sz 33           & M0   & M2   & M0$^{\mathrm{b}}$   & ISO-ChaI 224\\
 Sz 84           & M5   & M4   & M4.5$^{\mathrm{c}}$ & \\
 Sz 97           & M3.5 & M2   & M3$^{\mathrm{c}}$   & \\
 Sz 105          & M5   & M3.5 & M4$^{\mathrm{c}}$   & \\ 

\noalign{\smallskip}
\hline
\end{tabular}
\begin{list}{}{}
\item[$^{\mathrm{a}}$\citet{com98}]
\item[$^{\mathrm{b}}$\citet{gast92}]
\item[$^{\mathrm{c}}$\citet{hug94}]
\end{list}            
\end{table}                       

Table \ref{SP} gives spectral types for the ISOCAM detected candidate
low mass members of the Chamaeleon I dark cloud, 
obtained using the G5 telluric standards. In general, spectral types derived
using this set of standards for the known objects provide a slightly
better agreement with the optical types than those obtained from the
O8 spectral type set (see Table \ref{KNOWN}).
The values of $Q$, corresponding to the G5 telluric standards, for
each of our ISOCAM targets are listed in Table \ref{SP}.        
We also used the index $Q$ to derive spectral types for the three objects
with no mid-IR excess in our sample. We obtained M1 ($Q =$ 0.86) for
ISO-ChaI 158, M5 \hbox{($Q =$ 0.66)} for ISO-ChaI 238 and M0 ($Q =$ 0.93) for \hbox{ISO-ChaI 239.}
                                                                               
\begin{table}
\caption[]{Spectral type, veilings, and masses of the ISO candidates} \label{SP}
\begin{tabular}{lccccc}
Name &  $Q$   & Spectral Type$^*$ & r$_\mathrm{K}$ & T$_\mathrm{eff}$$^{**}$ & M/M\sun \\
\hline
\noalign{\smallskip}
ISO-ChaI 79  & 0.76 & M3         &   $+$0.36  & 3260 &  0.22  \\
ISO-ChaI 95  & 0.51 & M8(M6-7)   &   $-$0.07  & 2430 &  0.025 \\
ISO-ChaI 98  & 0.65 & M5         &   $-$0.14  & 2928 &  0.14  \\
ISO-ChaI 111 & 0.62 & M6(M5)     &   $-$0.13  & 2762 &  0.09  \\
ISO-ChaI 138 & 0.63 & M5.5       &   $-$0.24  & 2845 &  0.09  \\
ISO-ChaI 143 & 0.61 & M6.5(M5.5) &   $-$0.16  & 2679 &  0.085 \\
ISO-ChaI 154 & 0.92 & M0     &   $+$0.03  & 3850 &  0.60       \\
ISO-ChaI 209 & 0.90 & M1     &   $+$0.21  & 3720 &  0.50       \\
ISO-ChaI 220 & 0.69 & M4.5   &   $-$0.00  & 3011 &  0.12  \\
ISO-ChaI 224 & 0.92 & M0     &   $+$0.19  & 3850 &  0.45   \\
ISO-ChaI 225 & 0.83 & M2     &   $+$0.68  & 3426 &  0.30  \\
ISO-ChaI 235 & 0.62 & M6(M5) &   $-$0.09  & 2762 &  0.09  \\
ISO-ChaI 250 & 0.62 & M6(M5) &   $-$0.18  & 2762 &  0.09  \\
ISO-ChaI 252 & 0.82 & M2     &   $-$0.06  & 3426 &  0.30  \\
ISO-ChaI 256 & 0.65 & M5     &   $-$0.04  & 2928 &  0.12  \\
ISO-ChaI 282 & 0.73 & M3.5   &   $+$0.07  & 3177 &  0.19  \\
\noalign{\smallskip}
\hline
\end{tabular}

\vskip 0.2in

\noindent
$^*$Note: Spectral types between brackets are corrected by
the different water absorption intensity between pre-main 
sequence and main sequence stars of the same T$_\mathrm{eff}$, as 
estimated by \cite{luri99} for spectral types later than M6.
\vskip 0.2in

\noindent
$^{**}$Note: We adopt \citet{wil99} T$_\mathrm{eff}$ calibration that
comprises M sub-types later than M2. To assign T$_\mathrm{eff}$ to 
the three objects (ISO-ChaI 154, ISO-ChaI 209, and ISO-ChaI 224) in this
table with spectral types ealier than M2 we use \citet{keha95}
temperature calibration.                                            
\end{table}                                     

\citet{com00} proposed a second reddening-free $\mathrm{H_2O}$ vapor index, 
I$_{\mathrm{H_2O}}$, that compares the intensity of the absorption bands near 1.9 $\mu$m.
This index is defined by the relative fluxes in four narrow band filters
(0.05 $\mu$m width) centered at \hbox{1.675 $\mu$m,} \hbox{1.750 $\mu$m,} 2.075 $\mu$m, 
and 2.25 $\mu$m, liying in regions of reasonably good atmospheric
transmission. On the contrary, \citet{wil99} index $Q$ comprises regions close
to the limit of the $K$-band window and thus may be affected by a poor telluric
correction of the data. Assuming the \citet{rile85} reddening law over the 
$HK$ band (A$_{\lambda}~ \propto~ \lambda^{-1.47}$) and denoting the fluxes at each
narrow band filter by $f1$, $f2$, $f3$, and $f4$, respectively, the index
I$_{\mathrm{H_2O}}$
is expressed by:

\begin{equation}
\mathrm{I}_{\mathrm{H_2O}} = (f1/f2)(f4/f3)^{0.76}.
\label{dos}
\end{equation}          

\begin{figure}
\centering
\includegraphics[width=14cm]{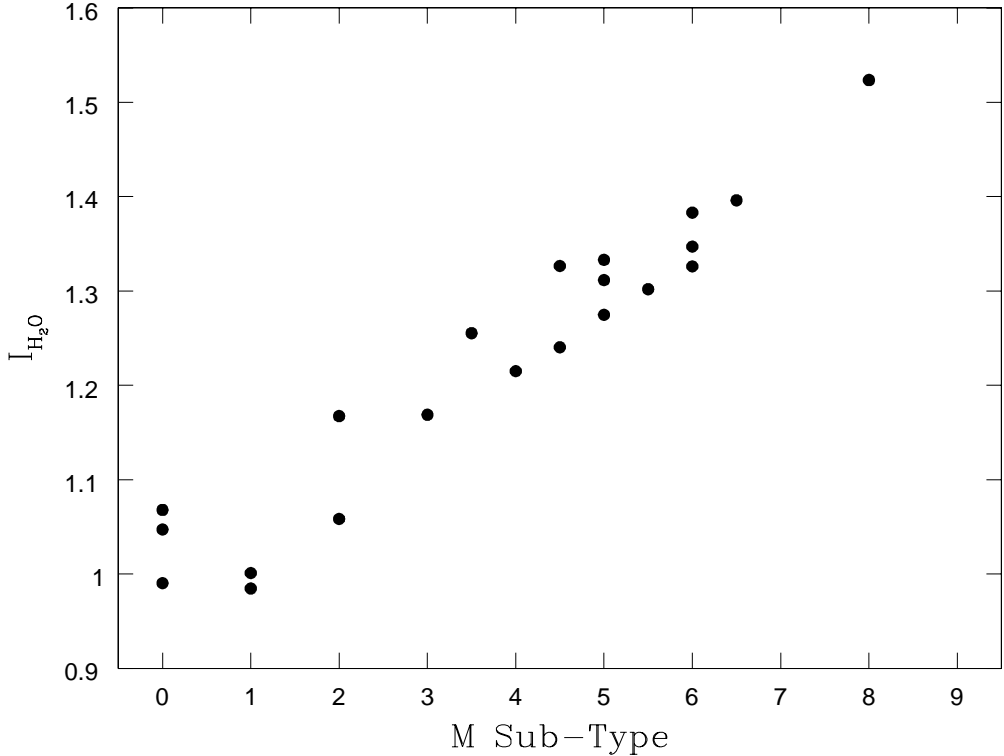}

\vskip -1.0in

\caption{The 1.9 $\mu$m water vapor index, I$_{\mathrm{H_2O}}$, defined by \citet{com00}
versus the M sub-types for all the stars observed with SOFI. The
spectral types are those obtained from the index $Q$, using the G5 spectral type stars as telluric
corrector, listed in Tables \ref{KNOWN} \& \ref{SP}. 
\hbox{ISO-ChaI 79} was not included in this figure as no reliable spectrum was obtained around
these wavelengths for this object (see Fig. \ref{Fig7}).}
\label{Fig8}
\end{figure}

We calculated this index for the all the stars observed, 
including the three previously known T Tauri stars (Sz 84, Sz 97, and
Sz 105) in the Lupus star-forming region 
and the 18 of the ISOCAM sources. We excluded ISO-ChaI 79 as no
reliable spectrum was obtained for this object around the wavelengths
of these narrow band filters (see Fig. \ref{Fig7}).
Fig. \ref{Fig8} shows the plot of the index I$_{\mathrm{H_2O}}$ versus
the M sub-types for these objects, derived from the index
$Q$ (see Tables \ref{KNOWN} \& \ref{SP}). A systematic trend is evident 
in the figure, although the correlation is rather poor. This trend 
gives some additional confidence to our spectral type estimations
based on the index $Q$ for the newly detected ISOCAM sources. 

We used the 6 objects in Table \ref{KNOWN} 
and the 1.9 $\mu$m water vapor index,
I$_{\mathrm{H_2O}}$, to obtain the following linear least-squares fit: 
\begin{equation}
\mathrm{M{sub\_type}} =  (-15.9 \pm 3.8) + (16.0 \pm 3.0) \times \mathrm{I_{H_2O}},
\label{tres}
\end{equation}     

with a correlation coefficient r $=$ 0.94. We estimate an uncertainty of
about 3 sub-classes in this calibration, considering typical errors in
our measurements of the index I$_{\mathrm{H_2O}}$ and the correlation coefficient
of Eq. \ref{tres}. This uncertainty is larger than that corresponding
to the index $Q$ calibration ($\sim$ 1.5 sub-classes). 
Relation \ref{tres} was derived from a the relatively small number of
observed stars with optically known spectral types. However this 
calibration allows us an initial confrontation between the water vapor
indexes. 

\begin{figure}
\centering
\includegraphics[width=14cm]{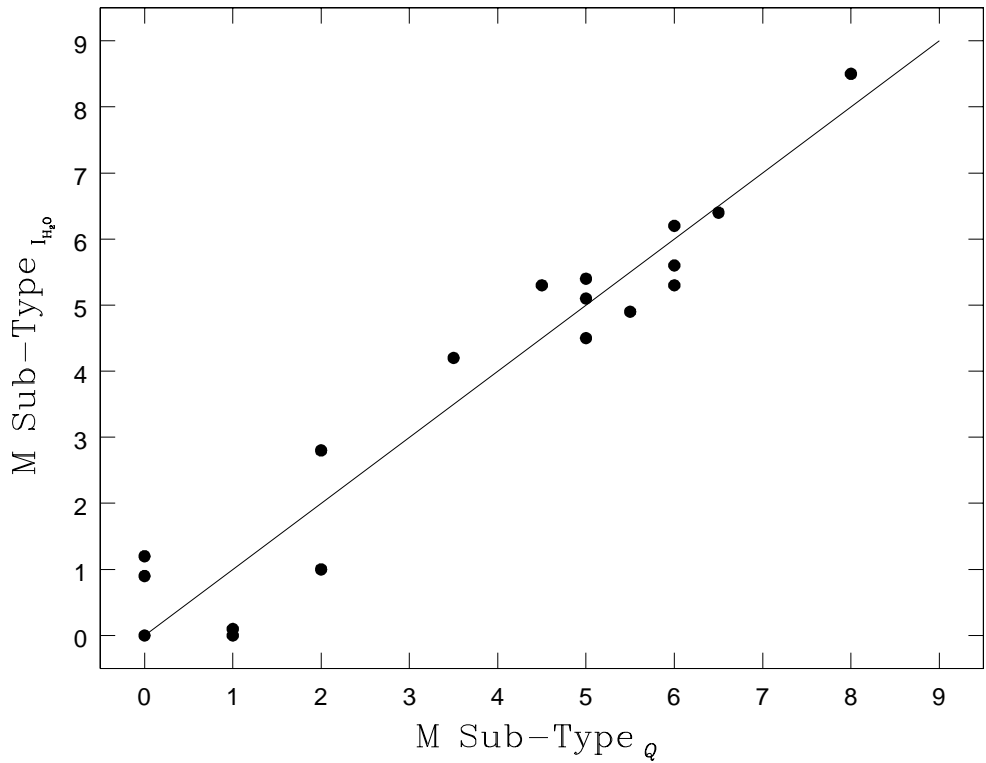}

\vskip -1.0in

\caption{Comparison between spectral types derived from the indexes
I$_{\mathrm{H_2O}}$ \citep{com00} and $Q$ \citep{wil99} for objects in Table \ref{ISOSP}. The solid
line corresponds to a linear relation with slope of 1, shown as reference. } 
\label{Fig9}
\end{figure}

In Fig. \ref{Fig9} we compare spectral types derived from both
the indexes $Q$ and I$_{\mathrm{H_2O}}$.  Spectral types for the ISOCAM selected
sources reported in this paper
agree within their relatively large uncertainties.
We caution, however, that additional
observations are necessary to
determine Eq. \ref{tres} more precisely
before a complete agreement between the two procedures can be claimed.

\cite{luri99} and \cite{com00} have suggested that the
water vapor indexes may saturate for stars as earlier as M6, 
losing sensitivity toward the latest M sub-types. The 
uncertainties in our index I$_{\mathrm{H_2O}}$ determination precludes us to distinguish
this effect.  A more accurate treatment is required to
estimate if the onset of water vapor saturation can significantly alter the
spectral type determinations. 

\subsubsection{Effects of veiling}  

As noticed by \citet{wil99} and \citet{cus00} the infrared excess produced by
the thermal emission from warm dust grains in circumstellar disks around young 
stellar objects may alter the index $Q$ and thus the spectral type estimations of our
targets.  To determine the $Q$ versus M sub-type relation these authors have
used main sequence stars of optically known M sub-types, essentially devoted
of any significant amount of infrared excesses.

This infrared excess emission can be characterized
by r$_\lambda = F_{\lambda}^\mathrm{exc}/F_{\lambda}^\mathrm{phot}$, where
$F_{\lambda}^\mathrm{exc}$ is the measured flux, including the excess (circumstellar) emission,
and $F_{\lambda}^{\mathrm{phot}}$ is the flux due to the underlying stellar photosphere. 
\citet{wil99} have simulated the spectra expected from young stars
surrounding by circumstellar matter, adding the
disk contribution to the spectra of known standard stars. 
These authors have then compared the index
$Q$ calculated for stars of known spectral types
{\it with} and {\it without} the disk emission. The spectral types obtained
from star+disk systems that have veiling effects of r$_\lambda \sim 0.2$ are about 1 sub-class 
earlier than those derived for objects without the disk contribution. This effect 
increases with r$_\lambda$ and can account for $\sim$ 3 sub-classes earlier
for r$_\lambda \sim 0.6$.

In Table \ref{SP} we give r$_\mathrm{K}$, the $K$-band excess, estimated for our target objects 
using the following relation, 

\begin{equation}
E(H-K) = (H-K) - (H-K)^{\mathrm{pho}} - 0.077A_\mathrm{V}
       = -2.5 \mathrm{log} [1/(1+r_\mathrm{K})] 
\label{cuatro}
\end{equation}           

\noindent
derived by \citet{mey97} \citep[see also][]{wil99, cus00}. In this expression
$(H-K)$ is the observed color (see Table \ref{ISOSP}) and $(H-K)^{\mathrm{pho}}$
corresponds to the intrinsic or photospheric color of each target. We have
adopted the calibration from \citet{wil99} to obtain intrinsic colors for our
targets. Only for three objects in Table \ref{SP} with spectral types earlier than M2
(ISO-ChaI 154, ISO-ChaI 209, and \hbox{ISO-ChaI 225)} we have obtained intrinsic colors
from \citet{keha95} as the \citet{wil99} calibration only comprises M2-9 stars.
Both color calibrations agree well in the common range. 
The third term in Eq. (\ref{cuatro}) corresponds to the excess in color due
only to the extinction to each individual target. $A_\mathrm{V}$ was derived
from $A_\mathrm{J}$ in Table \ref{ISOSP} according to \citet{rile85} redding law
($A_\mathrm{J} = 0.28 A_\mathrm{V}$). As noticed by \citet{wil99} and \citet{cus00} this
estimation provides only a lower limit to the veiling as Eq. (\ref{cuatro})
assumes r$_\mathrm{H} = r_\mathrm{J} = 0$ \citep[see also][]{mey97}.

The ISOCAM targets in Table \ref{SP} have r$_\mathrm{K}$ $<$ 0.2 with exception of
ISO-ChaI 79 and ISO-ChaI 225. We then expect our spectral type determinations using the $Q$
index be little affected by the effects of veiling except for these two
objects (ISO-ChaI 79 and ISO-ChaI 225), 
that may actually have spectral types of $\sim$ 2-3 sub-types later
than those given in Table \ref{SP}.  

\subsubsection{Surface gravity effects}

Pre-main sequence stars are known to have surface gravities
intermediate between those of dwarfs --luminosity class V-- and giants
--luminosity class III-- \citep[e.g.,][]{grme95, grla96}.
However we have determined spectral types for
our ISOCAM targets using the $Q$ index versus the M spectral type relation
derived for M type main sequence standards \citep{wil99, cus00}.
The water vapor telluric absorption in the 2 $\mu$m
spectral region is weaker in late-type giants than in dwarfs of the
same spectral type \citep[e.g.,][]{klha86}. For spectral types
later than M6 the water vapor absorption bands tend to be similar in depth both
for dwarfs and giants and then the spectral classification would result
quasi independent of the surface gravity. \citet{wil99} estimated a
systematic error of about 1-2 sub-classes earlier than
appropriate for stars earlier than M6 by assuming dwarfs rather than
giants gravities. 
 
For {\it pre}-main sequence stars \citet{luri99}
have found, however,  that the water absorption band is stronger than for
main sequence stars of the same T$_\mathrm{eff}$. These authors
have also found that spectral types derived from the $Q$ index
are $\sim$ 1 sub-class later for M6-M7 and $\sim$ 1--2 sub-classes
for $>$ M8 stars. This estimation is based on a comparison between
spectral types derived using the $Q$ index and optically determined
spectral types. We adopt this correction for the 5 objects
in Table \ref{SP} with spectral types later than M6. We indicate
the corrected spectral types between brackets in this table.
 
Lacking of a more appropriate calibration to estimate spectral types
for potential young stars, we estimate a {\it typical} uncertainty of about 1--2 sub-classes
in our spectral classification. This uncertainty is comparable to the expected
precision for the $Q$ index versus M-sub type relation as mention in
Sect. 3.3.1 \citep[see also][]{wil99}.

\subsubsection{Effective temperatures and masses}

To derive effective temperatures for our targets we have considered
and compared three calibrations from the literature: a) \citet{keha95},
b) \citet{wil99}, and c) \citet{luh99}. Both \citet{keha95} and \citet{wil99}
temperature scales are obtained using M dwarf stars.
\citet{keha95} calibration, combined with the \citet{bri98} determination,
extends from M0 to M8 spectral types, while \citet{wil99}
comprises M2-9 stars. The third scale, \citet{luh99}, provides
an intermediate scale between the luminosity class V and III objects, from M1 to M9.

The \citet{keha95} calibration is based on the \citet{scka82} work.
More recently \citet{bri98} used \citet{hen94} and \citet{kir93, kir95}
observations to test and continue this scale toward the latest M sub-types.
The \citet{wil99} calibration corresponds to a linear
least-squares fit to a single data set \citep{jon96}.
\citet{luh99} used \citet{leg96} data to obtain a temperature scale for 
dwarf stars. The model atmosphere of \citet{kir93} allows \citet{luh99}
to extrapolate the calibration from M6 to M9 \citep[see also][]{luri98}.
The temperature scale for the giants was derived from different data sets. 
In particular, \citet{van99} spectra were used to derive the fit for spectral types
earlier than M7 and \citet{per98} and \citet{ric98} works for giants in the M7-M9
spectral range.  Using these two independent calibrations, \citet{luh99} derived an
intermediate scale between the dwarfs and giants. 
 
Pre-main sequence objects have surface gravities intermediate between the
luminosity class V and III stars and thus \citet{luh99} calibration,
in principle, seems to be
the most appropriate. The numerical values of this calibration are
intermediate between those of \citet{keha95} and \citet{wil99} and, in
general, closer to \citet{keha95}'s temperatures. 

The application of \citet{luh99} and \citet{keha95} spectral types --
T$_\mathrm{eff}$ conversion
scales leads to results that are hard to conciliate with previous determinations
for the high mass members of the \hbox{Chamaeleon I} cloud. In particular some of the
stars with large near-IR excesses, when applying these calibrations,
have ages too old ($\sim$ 5 $\times$ 10$^7$ yr) in relation to the
$\sim$ 3-5 $\times$ 10$^6$ yr determined for the higher mass members of
the cloud by \cite{law96}.  On the contrary,
\citet{wil99}'s calibration provides consistent results, in 
in good concordance with previous works.  In addition this calibration is
based on an homogeneous set of standards whereas the other two involve
combinations of different data. \citet{luh99} calibration used \citet{kir93}'s
model to extrapolate to the latest M sub-types. Model atmospheres of M
dwarfs reasonably fit the overall shape of the observational data. However, several
difficulties in fitting more specific features still remain
\citep[see, for example,][]{kir93, all95}.  Thus we adopt the \citet{wil99}'s
calibration to place our targets on the HR diagram as it gives the clearest
results, based on an uniform scale. Only for three objects in Table \ref{SP} 
with spectral types earlier than M2 (ISO-ChaI 154, ISO-ChaI 209, ISO-ChaI 224)
we use the \cite{keha95} temperature scale. 
          
\begin{figure}
\centering
\includegraphics[width=18cm]{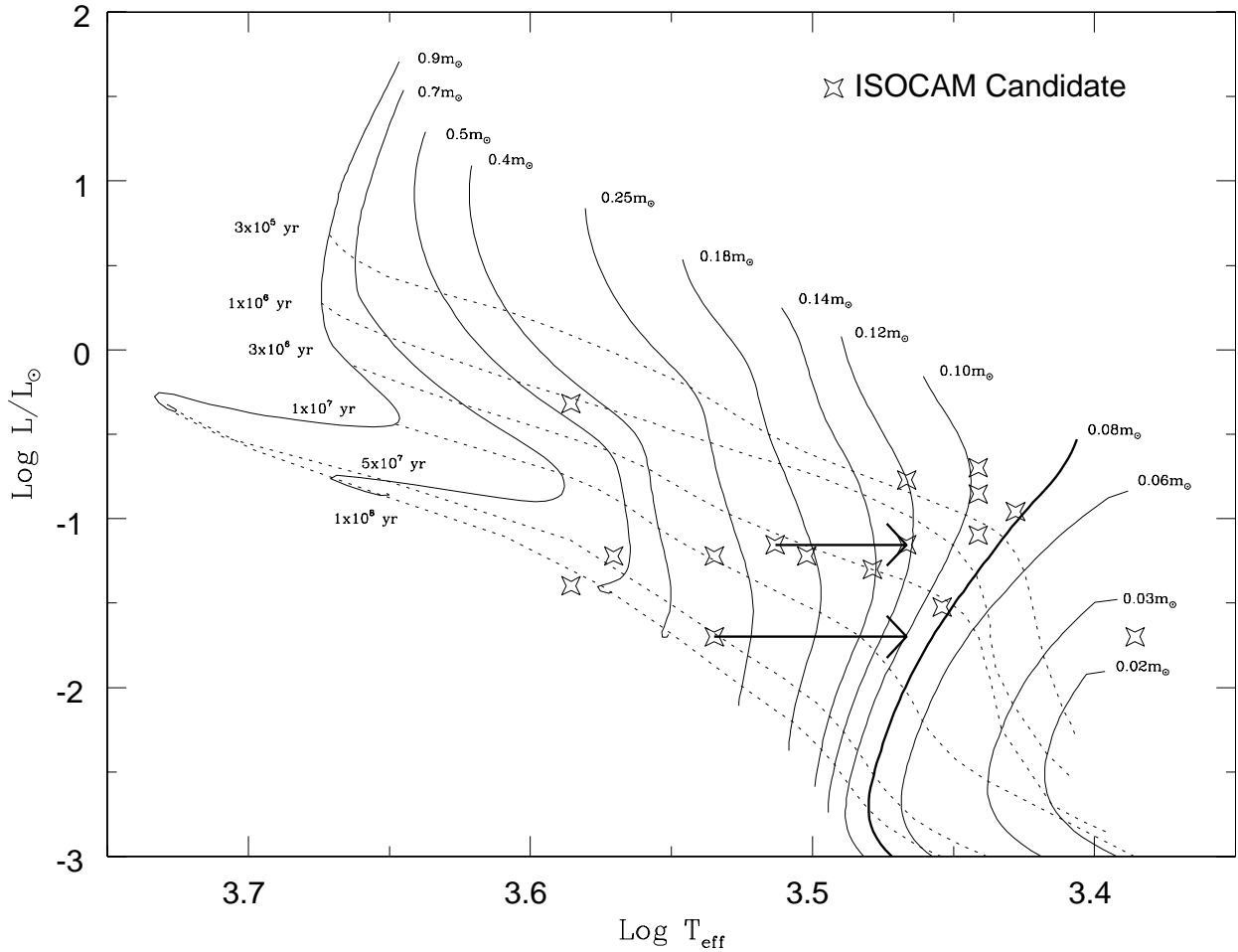}
\vskip -1.5in
\caption{HR diagram showing the positions of the ISOCAM selected
low mass stars. Stellar (not bolometric) luminosities were taken from
\citet{per00} (see also, Table \ref{ISOSP}) and
effective temperatures from Table \ref{SP}. We have adopted the
\citet{wil99} temperature calibration in this paper that comprises M2 to M9
spectral types. For the three objects in Table \ref{SP} with spectral types M0-1
(ISO-ChaI 154, ISO-ChaI 209, and ISO-ChaI 224)
we have used the \citet{keha95} temperature scale.
Pre-main sequence evolutionary tracks, indicated with continuous lines,  are
from D'Antona \& Mazzitelli (1998). The thick continuous line corresponds
to 0.08 M\sun, the H burning limit. The dashed lines correspond to the isocrones
calculated also by the same authors. The arrows indicate the displacements on
this diagram of ISO-ChaI 79 and ISO-ChaI 225, when correcting the corresponding
spectral types and effective temperatures due to the high values of veiling
(r$_\mathrm{K}$ $=$ $+$0.36 and $+$0.68, respectively) determined for these targets.
Accordingly, the lengths of the arrows show displacements of 2 and 3 sub-types
\citep[see Sect. 3.3.2 and][]{wil99}. The position of ISO-ChaI 79 roughly
coincides with ISO-ChaI 98, in this case.}
\label{Fig10}
\end{figure}

Fig. \ref{Fig10} shows the positions of the ISOCAM detected stars
in the HR diagram.  Luminosities have
been obtained from \citet{per00} (see also, Table \ref{ISOSP}). These authors have computed stellar 
(not bolometric) luminosities for the newly and previously known members
of the Chamaeleon I cloud and compared their luminosities determinations for the higher
mass pre-main sequence stars with those of \citet{law96}.  They have found a reasonable agreement
within a factor of 2.  Pre-main sequence evolutionary tracks
(continuous lines) and isocrones (dash lines) are from D'Antona \& Mazzitelli
(1998)\footnote{Available at http://www.mporzio.astro.it/~dantona/prems.html.}.

Several evolutionary tracks \citep[e.g.,][]{bur97, bar98, past99} that
are available, in principle, can be used to infer masses and ages. 
We have chosen D'Antona \& Mazzitelli (1998) model as these tracks cover
a wide range of masses, from 0.017 M\sun~ to 0.9 M\sun. Thus, they 
allow us to estimate masses and ages for all of our targets in a
homogeneous manner. In addition this set of evolutionary tracks and 
isocrones provide consistent and plausible results for our group of
targets, in reasonable agreement with the higher mass members of the
cloud. \citet{bar98} and \citet{past99} tracks do not comprise 
sub-stellar objects. \citet{bur97} model, on the contrary, provides
detailed calculations in the sub-stellar regime extending up to
0.2 M\sun.  The masses and ages derived from different tracks
and isocrones, in the common mass range and for a given T$_\mathrm{eff}$ 
calibration, typically agree within a factor of $<$ 2 in mass and $<$ 3 in age. 
This, however, introduces some uncertainty in the stellar/sub-stellar
nature of some of the ISOCAM detected objects analyzed here.
Table \ref{SP} lists masses for our targets derived
from D'Antona \& Mazzitelli's
tracks. 

The ISOCAM detected sources have sub-solar masses down to the H 
burning limit or even, in the case of ISO-ChaI 95 (Cha H$\alpha$ 1), below this limit. 
Three of the stars in Table \ref{SP} have previous estimations of masses.
For Sz 33 (ISO-ChaI 224) \citet{law96} derived masses of 0.3 M\sun~ and 0.4 M\sun,
from \citet{dama94} and \citet{swe94} evolutionary tracks, respectively. Our mass determination,
based on the \citet{keha95} temperature scale, agrees well with \citet{law96}'s estimation.

\citet{com00} estimated masses of 0.04--0.05 M\sun~ and 0.07--0.08 M\sun~ for
Cha H$\alpha$ 1 (ISO-ChaI 95) and Cha H$\alpha$ 2 (ISO-ChaI 111), respectively,
using \citet{bar98} and \citet{bur97} evolutionary tracks. Cha H$\alpha$ 1
is considered a bona-fide young brown dwarf member of the
cloud.  Cha H$\alpha$ 2 is a {\it transition} (sub-stellar/stellar) object with a mass close
to the H-burning limit. Considering typically errors in the mass determinations
(roughly a factor of 2, at this low limit) our estimations are consistent with
the sub-stellar or quasi sub-stellar nature of these objects and agree well
with previous works.

These targets span a range of ages from 1--3$\times$ 10$^5$ yr to
$\sim$ 10$^7$ yr. Only three objects (ISO-ChaI 225, ISO-ChaI 209, and ISO-ChaI 154)
in Fig. \ref{Fig9} have ages $>$ few $\times$ 10$^7$ yr.  
The median age of the rest of the newly detected sources,
is $\sim$ few $\times$ 10$^6$ yr. The range and median age of these stars
roughly correspond to the median age and to the spread in age found
by \citet{law96} for the higher mass members of this cloud.
                                      
Due to the high veiling of ISO-ChaI 225 (r$_\mathrm{K}$ $\sim$ $+$0.68) 
we may have underestimated the spectral type of this star in about
3 sub-type (see Sect. 3.3.2). Adopting a \hbox{T$_\mathrm{eff}$ $\sim$ 400 K}
cooler, this star shifts to the right on the HR diagram, corresponding
to decreasing values in both masses and ages (0.1 M\sun~ and 5 $\times$ 10$^6$ yr, 
respectively). In Fig. \ref{Fig10} we have indicated with arrows the displacements
that ISO-ChaI 225 as well as \hbox{ISO-ChaI 79} (r$_\mathrm{K}$ $\sim$ $+$0.36) 
experiment correcting the spectral types of these objects for their
high veiling values, according to \cite{wil99}'s estimation. In this case,
ISO-ChaI 79 practically coincides with the position of ISO-ChaI 98
(0.14 M\sun~ and 1-2 $\times$ 10$^6$ yr) on this diagram.

For ISO-ChaI 209 and ISO-ChaI 154 (spectral types M1 and M0,
respectively) we have used \cite{keha95} T$_\mathrm{eff}$ calibration, 
as \cite{wil99} relation comprises only spectral types later than M2.
The temperatures derived by \cite{wil99} are roughly 300 K cooler than
those obtained by \cite{keha95}, in the common range. If we adopt a
correction of this amount, these targets shift
to the right on the HR diagram, corresponding to decreasing masses
and ages (0.19 M\sun~ and 1 $\times$ 10$^7$ yr, for ISO-ChaI 209; 
0.4 M\sun~ and 3 $\times$ 10$^7$ yr, for ISO-ChaI 154), in better
agreement with the other ISOCAM detected sources.

\subsection{Detection of emission features in the ISO observed sample}

The ISOCAM detected sample analyzed here corresponds to a quite homogeneous
group. These thirteen sources have similar stellar luminosities and spectral indexes
(see Table \ref{ISOSP}) and comparable spectral types and masses (see Table
\ref{SP}). However, within our resolution and sensitivity limit, we detected
hydrogen emission lines in 5 of the 13 objects (i.e., 38\% of the sample, see Table \ref{ISOEW}).
These five ISOCAM objects are: ISO-ChaI 98, ISO-ChaI 143, ISO-ChaI 220, ISO-ChaI 252,
and ISO-ChaI 256 (see Figs. \ref{Fig4} \& \ref{Fig7}). We find no correlation
between the presence of H Paschen and Brackett lines in emission and the
near-IR color excesses or the amount of veiling estimated for each target.

\citet{nay96} and \citet{muz98} have investigated the formation of hydrogen
recombination emission lines in magnetospheric zones of the CTTS. In this model the
stellar magnetic field truncates the disk (at some inner radius) and the accretion
continues onto the central object following the magnetic field lines.
These authors suggest that, in general, emission lines are indicators of
disk accretion (infall) rather than stellar wind (outflow). Thus the ISOCAM
sources with H lines in emission are likely to be surrounding by
circumstellar disks.        
 
The homogeneity of the ISOCAM sample observed suggests that the lack of
H emission lines detection may be due to a projection or geometrical
effect. Sources with no emission lines are probably not favorably
oriented (i.e., practically edge-on disks). Higher resolution and sensitivity
data are required to properly test the effect of the disk inclination on the
H emission lines detection.               

\section{Summary and conclusions}
 
We obtained 0.95--2.5 $\mu$m moderate (R $\sim$ 500) resolution
spectra of 19 ISOCAM detected sources in the Chamaeleon I dark
cloud. Three of these objects (ISO-ChaI 158, ISO-ChaI 238, and ISO-ChaI 239), 
without near- and mid-IR excess, show no spectral characteristics typical of
young stellar objects (such as veiling or emission lines) and
are likely background M dwarfs.

Thirteen of the sources are candidate very low
mass members of the cloud proposed by \citet{per00} on basis
of the mid-IR color excess. The spectra of twelve of
these sources are relatively flat and featureless in this wavelength range.
The overall shapes and slopes of these spectra are typical of Class II young
stellar objects \citep{grla96}.  Both atomic and molecular lines (when in
absorption) are partially veiled (see Table \ref{ISOEW}), suggesting the presence
of continuum emission from circumstellar dust.
In addition some of the sources show Paschen and Brackett lines
in emission, also indicating the presence of disks.
\citet{per00} had already classified
these objects as potential Class II members of the Chamaeleon I 
dark cloud on basis of the infrared spectral index \hbox{($\alpha_{\mathrm{IR}}$ $=$ 
dlog($\lambda$F$_{\lambda}$/dlog($\lambda$)).} These objects have
$-$1.6 $<$ $\alpha_{\mathrm{IR}}$ $<$ 0.3, typical of Class II objects (see Table
\ref{ISOSP}), with exception of ISO-ChaI 250 ($\alpha_{\mathrm{IR}}$ $=$ $-$1.9),
classified as a ClassII-III object by \citet{per00}. 
ISO-ChaI 154 is the only source with mid-IR excess that shows 
near-IR spectral features characteristic of Class III objects or WTTS
\citep{grla96}.  

Three additional sources
(Sz 33, Cha H$\alpha$ 1, and Cha H$\alpha$ 2) are previously known
members of the cloud. In particular, Cha H$\alpha$ 1 and Cha H$\alpha$ 2
are a bona-fide young brown dwarf and a {\it transition} object (with a mass
close to the H-burning limit) optically detected by \citet{com98}.
We find no substantial difference between the spectra of these sub-stellar
or quasi sub-stellar objects and those of very low mass stars in our sample
at the resolution of our data.  We notice, however, that allowing for typical
uncertainties of about a factor of 2 in our mass determinations at least some of our
targets are probably {\it transition} objects that lie in the boundary region
between the stellar and the sub-stellar regimes.                     

We apply the 2 $\mu$m water vapor index defined by \citet{wil99} to estimate
spectral types for the 13 mid-IR excess sources observed. These stars have spectral
types M0--8. We use Persi et al.'s stellar luminosity determinations,
in combination with D'Antona \& Mazzitelli latest pre-main sequence
evolutionary tracks, to estimate masses and ages.
The ISOCAM detected mid-IR excess sources 
have sub-solar masses down to or, in the case of ISO-ChaI 95 (Cha H$\alpha$ 1),
even below the H-burning limit. These objects span a
wide range of ages from 1--3$\times$ 10$^5$ yr to $\sim$ $\times$ 10$^7$ yr.
The median age of the newly detected sources is $\sim$ few $\times$ 10$^6$ yr.  
The range of ages as well as the media age agree well with those
found by \citet{law96} for the higher mass members of this cloud.

ISO-ChaI 225 has a high veiling (r$_\mathrm{K}$ $\sim$ 0.68) and thus our spectral type
determination may be underestimated up to about 3 sub-types. Assigning a
spectral type 3 sub-type later (M5 instead of M2), this object
has an age consistent with the rest of the ISOCAM detected low mass
members of the Chamaeleon I dark cloud.  \hbox{ISO-ChaI 209} and \hbox{ISO-ChaI 154}
have spectral types M1 and M0, respectively. We used \cite{keha95} calibration
to assign T$_\mathrm{eff}$ to these targets. This temperature scale is
$\sim$ 300 K hotter than \cite{wil99}'s scale. Considering this
correction both stars shift to the right on the HR diagram and have
ages $\sim$ 1-3 $\times$ 10$^7$ yr, in reasonable agreement with the
rest of the ISOCAM detected sources and higher mass members of the
cloud. 

We found that 12 of the 13 ISOCAM detected candidate members of the \hbox{Chamaeleon I}
dark cloud show near-IR spectroscopic features typical of Class II objects.
Additional spectroscopic observations are required to investigate the
physical properties of the 21 remaining very low luminosity candidate members
of the cloud proposed by \citet{per00}. This group of objects may include about 19
stars with similar near-IR spectroscopic characteristics as Class II objects in
view of the high detection rate of likely young members of the cloud found
in this contribution. Near-IR spectroscopy is required to estimate masses
for these 21 remaining objects. Combined mass determinations for all the
ISOCAM detected objects would allow us to explore the behavior of the IMF
of the cloud in the sub-stellar regime. This function is presently very
poorly known for stars close and below the H burning limit.

\begin{acknowledgements}
 
We are grateful to the ESO staff for assistance
during the observing run, specially to Leonardo
Vanzi for technical support with SOFI. We thank the referee, Thomas 
Greene, for helpful criticisms and suggestions that improved the 
content and presentation of this paper. 
This research has made use of the SIMBAD database,
operated at CDS, Strasbourg, France.
 
\end{acknowledgements}

\end{document}